\documentclass[fleqn,3p,times, preprint]{elsarticle}
\usepackage[utf8]{inputenc}
\usepackage[T1]{fontenc}
\usepackage{subcaption}
\usepackage{ulem}
\usepackage{ecrc}
\usepackage{amssymb}
\usepackage{amsmath}
\usepackage{xcolor}
\usepackage{comment}

\volume{}
\firstpage{1}
\journalname{}
\runauth{Georgiou et al.}

\begin{document}
\begin{frontmatter}
\dochead{}
\title{Retroactive Interference Model of Power-Law Forgetting}

\author[aff1]{Antonios Georgiou}
\author[aff1]{Mikhail Katkov}
\author[aff1]{Misha Tsodyks\corref{cor1} }
\cortext[cor1]{ Corresponding Author \ead{mtsodyks@gmail.com} }
\address[aff1]{Weizmann Institute of Science, Department of Neurobiology, Rehovot, 7610001, Israel}

\begin{abstract}
Memory and forgetting constitute two sides of the same coin, and although the first 
has been rigorously investigated, the latter is often overlooked. A number of experiments under the realm of psychology and
experimental neuroscience have described the properties of forgetting in humans and animals, showing that forgetting exhibits a power-law relationship with time. These results indicate a counter-intuitive property of forgetting, namely that old memories are more stable than younger ones. We have devised a phenomenological model that is based on the principle of retroactive interference, driven by a multi-dimensional valence measure for acquired memories. The model has only one free integer parameter and can be solved analytically. We performed recognition experiments with long streams of words were performed, resulting in a good match to a five-dimensional version of the model.   
\end{abstract}

\end{frontmatter}


\thispagestyle{empty}

\section*{Introduction}

Memory has been often associated solely with the property of persistence, that is the ability to retain and retrieve information with the passage of time. Another equally important characteristic of memory however, is transience, or in other words, the ability to forget and discard information that could be no longer relevant. This process is considered crucial for memory and it is hypothesized to be essential for adaptive behavior \cite{Richards2017}. Traditionally, since Ebbinghaus`s seminal study \cite{Ebbinghaus1885}, forgetting has been described with the use of the retention curve. This curve is a continuous function of time $R(\tau)$, which denotes the probability that a memory of age $\tau$ still exists (i.e not yet forgotten). The shape of the retaining function has been investigated through the examination of experimental data over the last century, with some of the authors proposing universal power-law decay \cite{Wixted1990,Wixted1991,Kahana2002} (but see \cite{Fisher2018}): 
\begin{equation}\label{eq:retention_prop}
R(\tau)\propto \tau^{-\alpha}
\end{equation}
Conversely, we can define a \textit{forgetting} rate function $F(\tau)$, expressing the probability that an available memory of age $\tau$, will be forgotten within the next time interval $dt$. The two functions are related under the equation:
\begin{equation}\label{eq:forgetting}
R(\tau)\approx\prod_{t=0}^{t=\tau/dt}(1-F(t)dt) \approx e^{-\int_{0}^{\tau}F(t)dt}
\end{equation}
which simply expresses the fact that in order to remain for time $\tau$, the memory has to remain for each of the intervening time intervals. Equation \ref{eq:forgetting} can be inverted to
\begin{equation}\label{eq:2}
F(\tau)=-\frac{R'(\tau)}{R(\tau)}
\end{equation}
This equation means that while the retention function is monotonically decreasing with time (if we assume that extinguished memories cannot be reinstated), the forgetting function could in principle be both decreasing and increasing, depending on the decay speed of the retention function. In particular, an exponential retention function is a borderline case which results in a forgetting function that is independent of time, i.e. all memories have the same probability to be forgotten, irrespective of their age.  

Substituting \eqref{eq:retention_prop} into \eqref{eq:2}, we get that for power-law decay of retention, the forgetting rate will decay in time at an inversely proportional manner, regardless of the value of the exponent $\alpha$:
\begin{equation}\label{eq:4}
F(\tau)\propto\frac{\alpha}{\tau}
\end{equation}
In other words, memories that are older are more resilient (have lower probability to be forgotten at any given moment). Most neural network models of memory however, treat forgetting as a process that is linear in time, i.e the older the memory, the more probable it is to be forgotten at any given moment, which is the core idea of forgetful learning and memory palimpsests \cite{Nadal1986}.

The interest in mathematical forms of memory curves was encouraged by the hope that they may shed light int the mechanisms of remembering and forgetting. While this is not necessarily true, as different mechanisms could result in similar mathematical forms of forgetting (see below), memory models should still be at least broadly compatible with observations, and ideally provide some insights into their computational advantages to potentially alternative schemes. Mechanisms that are usually considered in relation to forgetting are passive decay of memories, interference and consolidation (see e.g. \cite{Wixted2004-R}). Decay theories state that memories are degraded with time, and are completely forgotten when a threshold is reached. On the other hand, the more popular interference theories suggest that prior (proactive) or subsequent (retroactive) learning disrupts memory consolidation and therefore memories are forgotten (For a review of both cases see \cite{Wixted2004-R}). A simple and elegant mathematical model of the first type is presented in \cite{Kahana2002}. While it would appear that passive decay of memory strength should result in new memories gradually replacing the older ones, Kahana and Adler showed that when new memories are characterized by variable initial strengths and decay rates, and are forgotten when the strength dips below threshold, retention function converges to $1/\tau$ scaling in the limit of large $\tau$. Importantly, the necessary condition for this property is that the distribution of decay rates extends all the way towards zero, i.e. some memories don't decay with time. For example, in the simplest case when memory decay is characterized by a linear function: $S(t) = a - bt$ with random positive $a$ and $b$, one can show that asymptotic scaling for the probability that a memory is still available at time $\tau$ after acquisition is given by
\begin{equation}
    R(\tau) \approx \frac{P_b(0)}{\tau}<a>,
\end{equation}
where $P_b(b)$ is the probability density of the decay rate $b$ and $<a>$ is the average value of the initial memory strength (see Appendix for a derivation). The condition that $P_b(0)>0$ also means that the average life-span of a memory is infinite. We conclude that while this study to some extent demystifies the power-law scaling of retention curves, the assumption about the passive decay of memories seems to contradict the well-documented effect of memory interference \cite{Wixted2004} and the model does not provide any mechanistic reasons for why some memories decay with time and some not, which is a crucial requirement for power-law forgetting. An alternative model that combines passive decay and interference was proposed in \cite{Lewandowsky}, where memories are characterized by a ratio of time since their acquisition to that of other memories. Recall probability is assumed to depend on its 'distinctiveness', defined as an inverse of acquisition time ratios averaged over all other memories . On one hand, interference is involved since different memories interact to determine their distinctiveness; on the other hand, when time passes without any new memories being acquired, distinctiveness of all memories, and hence their recollection, declines, indicating that passive decay is also effectively at play. The authors show that this mathematical model accounts for experimental retention curves as well as other well-known phenomena in recall literature, such as recency-to-primacy gradient. However, the model rests on several strong assumptions, e.g. the time since acquisition of each memory has to somehow be encoded, and a particular mathematical form of distinctiveness measure has no mechanistic underpinning.   

In the current contribution, we aim at a forgetting mechanism that would be compatible with realistic retention curves, contain as few assumptions as possible, and could have a clear functional interpretation. To this end, we propose a phenomenological model that parallels the concept of retroactive interference and captures the statistical properties of forgetting that were previously discussed. Similar to \cite{Kahana2002} it simplifies the memory retention as a binary process (available/forgotten) and introduces the notion of memory strength (that is generalized to be a multi-dimensional valence). However it does not assume an independent and passive decay of memory valence, rather it proposes a specific process of forgetting based on the idea of retroactive interference. The process proposed has a clear functional interpretation of trying to keep important memories while discarding less important ones. Besides valence dimensionality, the model has no free parameters to tune and results in an asymptotic scaling of retention function similar to that of \cite{Kahana2002} without any additional assumptions. Furthermore, we performed a series of recognition experiments designed to quantitatively test the predictions of the model and found that it provides a good match to the data. 

\section*{The Model}

\begin{figure}
    \centering
    \includegraphics[width=0.7\linewidth]{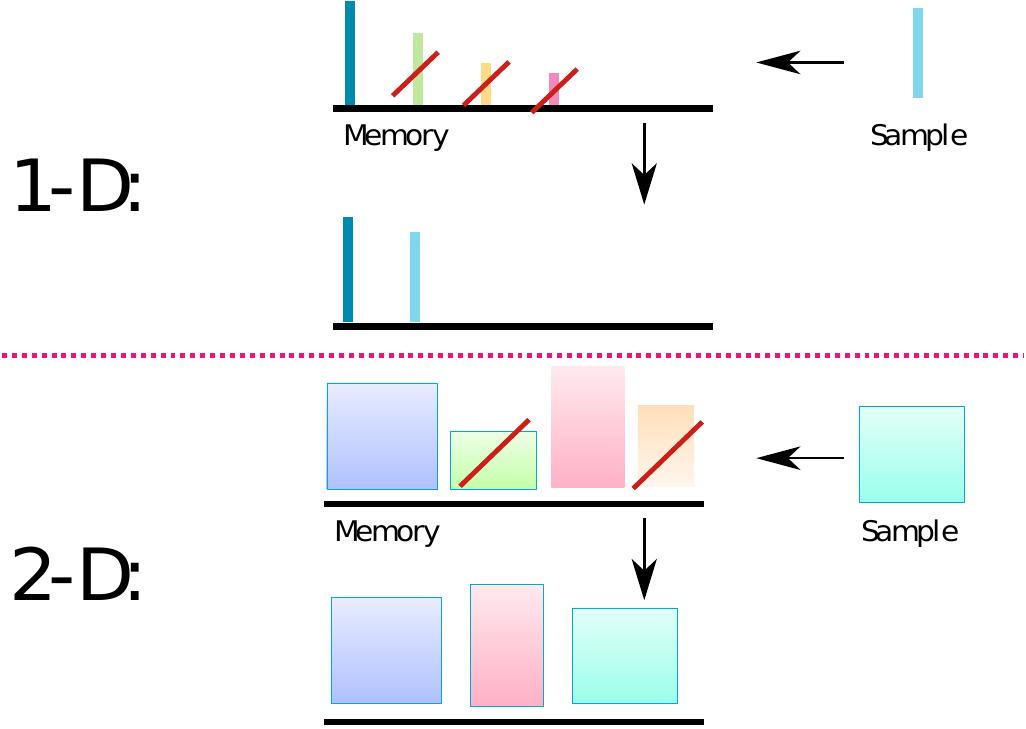}
    \caption{ {\bf Interference model of forgetting.} {\bf1-D.} Each item is represented as a thin vertical bar. The height of the bar corresponds to the valence of an item. The top row bars above the black line represent items that are stored in memory just before the acquisition of a new item, shown on the right (Sample). All the items that have smaller valence (bar height) than the new item are discarded from memory (crossed by red bar), and the new item is added. Bottom row represents the memory content, after the new memory is acquired. {\bf2-D.} Same as 1-D, but each memory item has 2 valences, represented by the width and height of a rectangular. In this case, all the items that have both valences smaller than the corresponding valences of the new item are discarded. }
    \label{fig:cartoon}
\end{figure}

To illustrate the main idea of our model, let us first consider a system that continuously acquires new memory items, each characterized by the scalar value $v$ (valence), considered to be a measure of its importance and independently sampled from a distribution $P(v)$. The form of this distribution can be arbitrary, but we assume that it is not changing with time. For simplicity, we assume that memories are acquired at a constant rate (one new memory per time unit). Each time a new item is sampled, all the previously stored items that have a smaller valence are discarded ('forgotten'; see Fig. \ref{fig:cartoon}, upper panel). Therefore, the total number of stored items will increase if items with relatively small valences are sampled but can suddenly decrease if the sampled element is very potent. This process can be regarded as a crude approximation to retroactive interference. By the construction of the model, at any given moment the valence of the stored units will be a monotonically increasing function of their age, since units are retained only if following units have a smaller valence and are discarded otherwise. Therefore, the probability that a unit will be forgotten at the next time step is strictly a decaying function of its age, i.e. one of the most counter-intuitive features of memory retention is inherently captured by the model. Mathematically, the retention function $R(t)$ is defined as the probability that a memory item is still retained in memory $t$ time steps after its acquisition, and can be computed as:
\begin{equation}\label{eq:5}
R(t)=\int_{-\infty}^{\infty}dv P(v)\bigg[1-\int_{v}^{\infty}dv'P(v')\bigg]^{t}
\end{equation} 
where the term in square brackets expresses the chance that an item with valence $v$ survives the acquisition of one extra item. With the change of variables $v\rightarrow y(v) = \int_{v}^{\infty}dv' P(v')$, equation (\ref{eq:5}) converts to:
\begin{equation}\label{eq:6}
R(t)=\int_{0}^{1}(1-y)^{t} dy=\frac{1}{t+1}
\end{equation}
We see that this simple model exhibits the uniform power-law scaling of memory retention for all times. Importantly, there are no free parameters that affect the retention properties of the model, in particular the form of the probability distribution of valences, $P(v)$ has no effect on the model behavior. The $1/t$ scaling of the retention curve implies that the average number of memories does not saturate with time but continues to grow, which is an attractive feature of the model. However, if we compute the average number of items in memory after a long time $T$ from the beginning of the acquisition process, we obtain
\begin{equation}\label{eq:7}
N(T)=\sum_{t=1}^{T}\frac{1}{t+1}\approx \log (T)
\end{equation}
i.e. the number of stored items is very small in relation to the total number of sampled ones. In particular, even after $T=10^8$ time units (several years of learning if one assumes a new memory acquisition per second), no more than twenty memories are retained, which is clearly not a reasonable estimate. Furthermore, the assumption that there is a single metric of importance is also unrealistic. Each piece of acquired information might be very important in one context but trivial in another (see also Discussion below). This idea can be easily translated into the model by introducing a multidimensional valence distribution, where each component of the sample $\textbf{v}$ represents its valence on a different domain. The forgetting rule in this case is expanded to all dimensions, and for an item to be forgotten, it is required that the newly acquired sample has a larger valence value on all axes (see \ref{fig:cartoon}, lower panel, for 2-dimensional case). The retention function in this extended model cannot be expressed in a closed form but can be iteratively computed with the following scheme:
\begin{equation}
\label{eq:recursive}
    R_n(t) = \frac{1}{t+1}\sum_{k=0}^t R_{n-1}(k)
\end{equation}
with n being the number of dimensions and $R_1(t)$ being the retention curve of the 1-dimensional model (equation \ref{eq:6} above, for detailed derivation see \ref{ap:1}). Repeated application of equation \ref{eq:recursive} allows the exact calculation of the retention curve for arbitrary $n$. Assuming large t, this expression approximates to
\begin{equation}
    R_n(t) \approx \frac{1}{(n-1)!}\frac{\log ^{n-1}(t+1)}{(t+1)},
    \label{eq:ret}
\end{equation}
which has the same scaling as in the one-dimensional case \eqref{eq:6} with a logarithmic correction. This correction aggregated over a long time $T$, leads to the total number of retained memories given by
\begin{equation}
   N_n(T)  \ =\sum\limits_{t=1}^T R_n(t)  \approx \frac{1}{n!}\log ^n(T).
\end{equation}

\begin{figure}[ht]
    \centering
    \includegraphics[width = \linewidth]{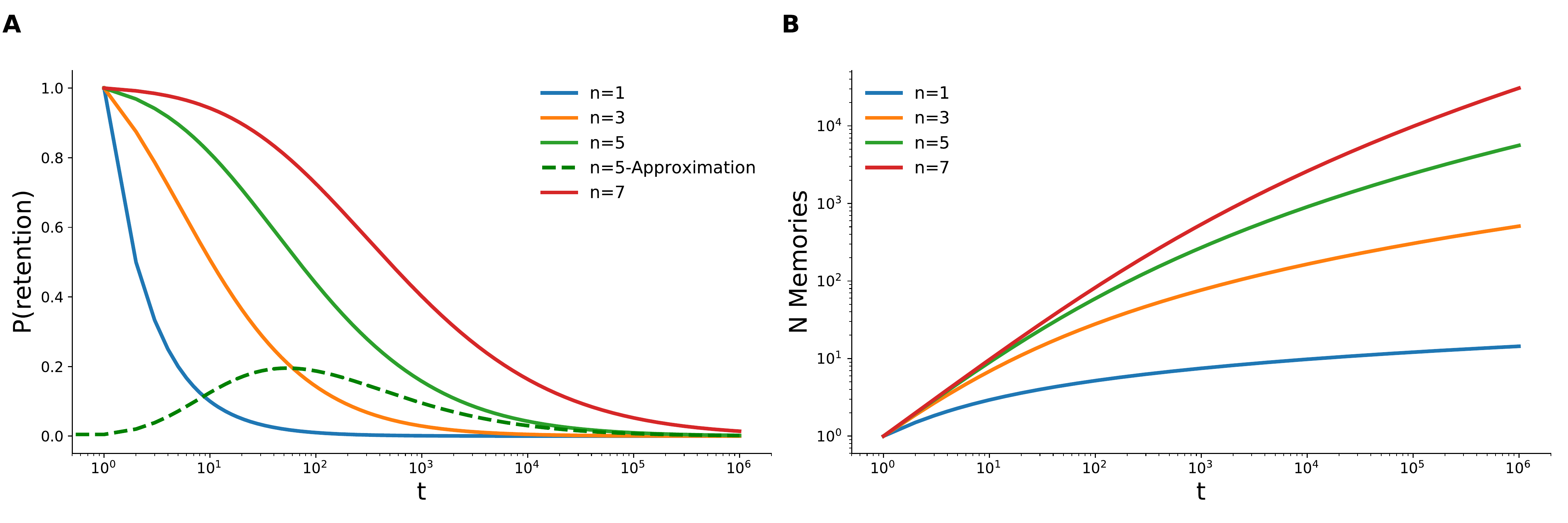}
\caption{ {\bf Theoretical results.} {\bf A.} Theoretical retention curves \eqref{eq:recursive} for different number of valence dimensions $n$. The dashed green line shows the asymptotic approximation of equation \eqref{eq:ret} for $n=5$ that converges to an exact curve from $T \approx 10^4$, {\bf B.} The average number of retained memories accumulated as a function of elapsed time from the beginning of acquisition.}
\label{fig:theory}
\end{figure}
Figure~\ref{fig:theory} shows the plots for $R(t)$ and $N(T)$ for several values of dimensionality. For example, we see that for $n=5$ the number of retained memories after $T=10^8$ steps of acquisition is around few tens of thousands, which appears to be a reasonable estimate \cite{Landauer1986}.

The above analysis shows that in the multidimensional case, the retention curves deviate from simple power-law functions due to logarithmic corrections. One can still approximate the retention curve with a power-law function with a slowly changing exponent:
\begin{equation}
    R_n(t) \approx c(t) t^{\alpha (t)}
\end{equation}
where the exponent $\alpha(t)$ can be estimated as 
\begin{equation}
    \alpha(t) = \frac{d(log(R_n(t))}{d(log(t))}
\end{equation}
(see Fig. \ref{fig:theory}). One can see that the scaling exponent is slowly reduced to $-1$ for very large times, remaining significantly above that asymptotic value even for times as large as $10^8$. The asymptotic expression of the exponent can be derived by the asymptotic expression for the retention curve, Eq. \ref{eq:ret}, resulting in 
\begin{equation}
    \alpha(t) \approx -1 + \frac{n-1}{log(t)}.
    \label{eq:alpha_approx}
\end{equation}

\begin{figure}
    \centering
    \includegraphics[width=\linewidth]{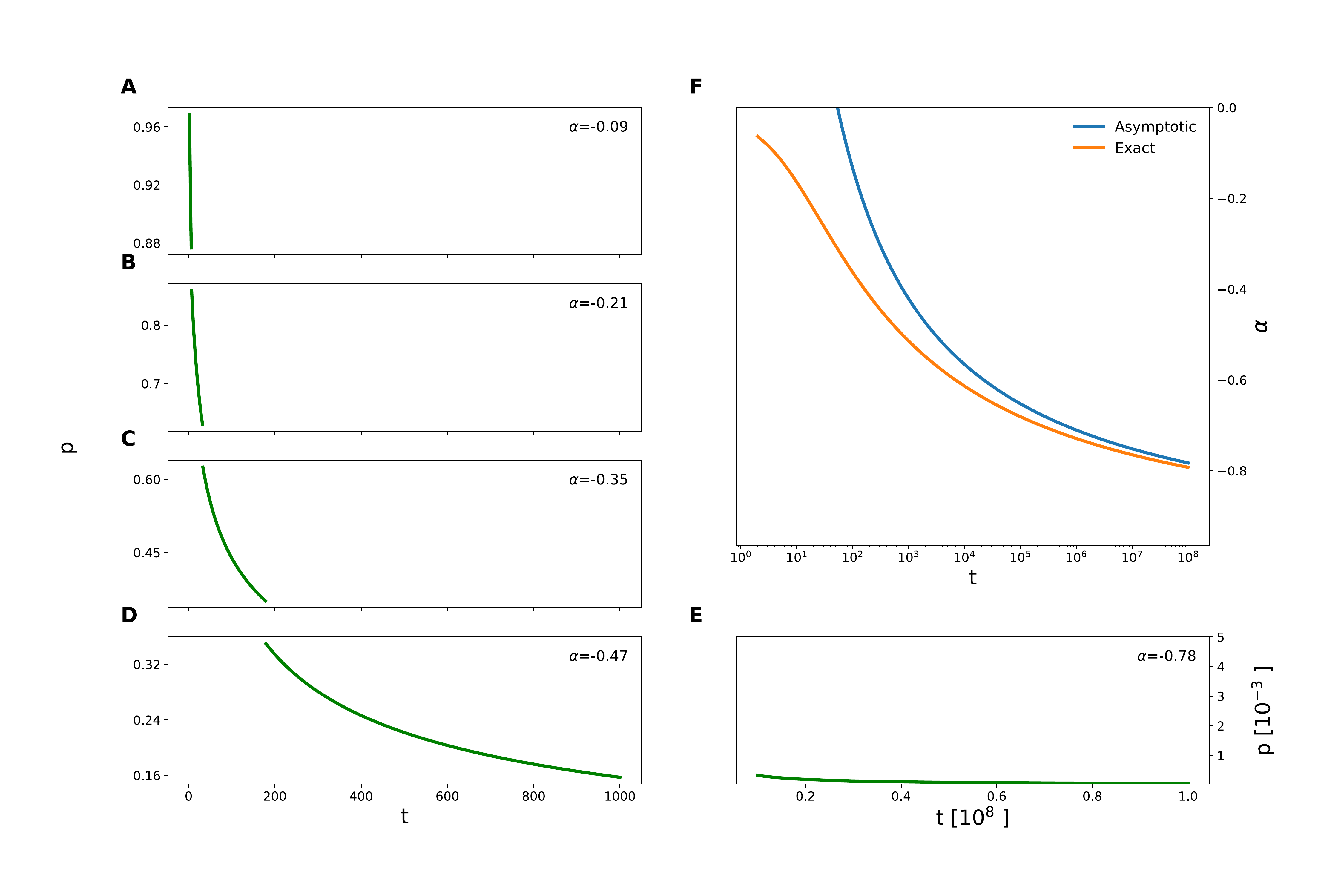}
    \caption{{\bf Power fit of theoretical retention curves.} {\bf A-E} Theoretical retention curve computed with equation \eqref{eq:recursive} for $n=5$, plotted for different time windows. In the inset, the estimated value of the power $\alpha$ for the corresponding window is shown. {\bf F} The dependence of $\alpha$ on time (orange curve). The value of $\alpha$ very slowly approaches $-1$, such that even for $T=10^8$ it is still about $-0.8$. For comparison the asymptotic estimate of $\alpha$ given by equation \eqref{eq:alpha_approx} is shown in blue. }
    \label{fig:alpha}
\end{figure}

\section*{Experiment}

\begin{figure}
    \centering
    \includegraphics{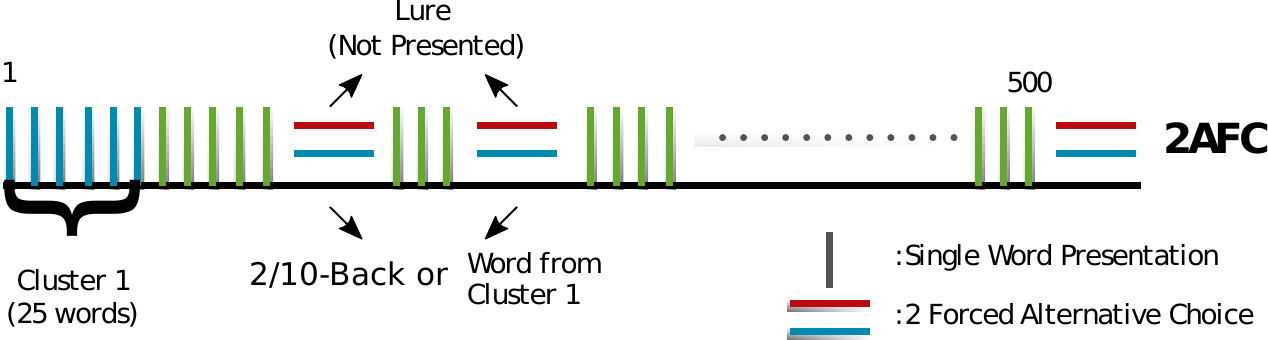}
    \caption{ {\bf Experimental protocol.} A series of vertical bars represents word presentations. Pairs of horizontal bars represent a delayed recognition task, where participants were presented with one word shown to them previously and one lure word. Participants were requested to click on the word they felt that they saw before. In total 500 words were presented and all first 25 words were queried at different moments. Additionally, participants undertook recognition tasks for the second (25) and tenth (25) back word from the time of inquiry resulting in a total of 75 tasks per participant. The 2/10 back conditions were conducted in blocks, i.e no question from the first 25 words was asked between presentation of the item to be queried and that query itself.  }
    \label{fig:exp_design}
\end{figure}

\begin{figure}
\begin{subfigure}{.0\textwidth}
\captionsetup{labelformat=empty} 
\caption{} 
\label{fig:avg_ret_all}
\end{subfigure}
\begin{subfigure}{.0\textwidth}
\captionsetup{labelformat=empty} 
\caption{} 
\label{fig:avg_ret_filt}
\end{subfigure}
    \centering
    \includegraphics[width=\linewidth]{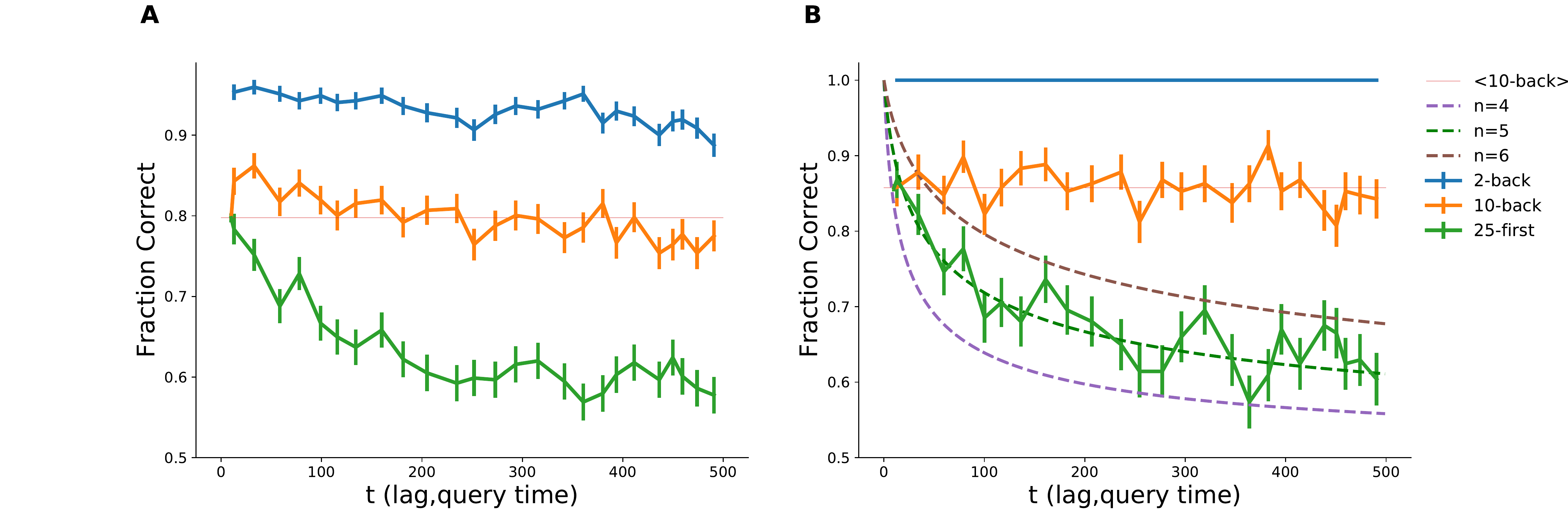}
    \caption{ {\bf Experimental results.} {\bf A} Recognition performance for all participants who passed the qualification task (see Methods). {\bf B} Recognition performance for participants that were perfect in the 2-back task. The experimental retention curve (recognition performance vs the lag between word presentation and delayed recognition task positions, solid green curve) declines for both groups. One can observe that performance in the  2-back task (solid blue curves, time here indicates the position of the recognition task) is declining with time, It is generally assumed that working memory capacity is 3-5 items, therefore the drop of performance in the 2-back task indicates that a progressively smaller fraction of words was acquired towards the end of the trial, due to fatigue, loss of attention, or other reasons. Selecting the group of participants (in panel {\bf B}) that are perfect in the 2-back task we ensured that all the words were acquired during presentation. For this group of participants, the performance in the 10-back task (orange solid curves, time indicates query position) remains constant throughout the experiment (compare with faint orange line representing mean performance in the 10-back task). The dashed lines represent theoretical retention curves computed with equation \ref{eq:retention} corrected for guessing (see Methods), for different number of dimensions $n$. }
    \label{fig:exp}
\end{figure}

To test whether the model conforms with human memory performance and to estimate the number of dimensions for the valence distribution of memory, we designed an experimental protocol based on the two-alternative-forced-choice delayed recognition task \cite{Standing1973}. The experiment was performed on Amazon's Mechanical Turk\textregistered $ $ platform. Participants were presented with a sequence of $500$ words, intermittent with recognition attempts. During recognition inquiries participants were prompted to select between two words on the basis of which word they remembered as having previously appeared: one choice constituted a word presented earlier in the sequence (either 2 or 10 words before the recognition attempt or one of the first 25 presented) and another one a lure (see Figure~\ref{fig:exp_design}). Following \cite{Standing1973} we make a simplifying assumption that if a previously presented word is still in memory, the participant will provide a correct answer, otherwise the response is going to stem from guessing. The experimental results are shown in Figure~\ref{fig:exp}. Figure~\ref{fig:avg_ret_all} shows the results for all 471 participants in the experiment. One can observe that the probability to recognize a word decays towards chance level (50\% correct recognitions) as a function of lag between presentation and inquiry (green line). The probability of recognizing the word presented 10 (10-back task) or 2 (2-back task) positions before the recognition prompt also declines as the experiment furthers in time (blue and orange lines). This could result from either proactive interference (when previously memorized items interfere with an acquisition of new words), or general fatigue accompanied with diminished attention, leading to disruption of new word acquisition. Since the short-term memory capacity is estimated to be 3--5 items \cite{Cowan2007}, we conjectured that the last 2 words, if acquired, should stay in short-term memory. We therefore selected 197 participants who exhibited perfect performance on the 2-back recognition task (see \ref{fig:avg_ret_filt}). Indeed, these participants show no decline of performance for the 10-back test either, indicating the absence of forward interference. Their retention performance (green line) is in agreement with the theoretical prediction for $n=5$ (dashed green line). 

\section*{Discussion}
We proposed a simple phenomenological model of forgetting that is broadly compatible with power-law retention curves reported in the earlier literature and with focused recognition experiments performed specifically for this study. In particular, the model results in power-law retention curves with exponents that very slowly decline towards $-1$, remaining significantly above this asymptotic value for all realistic time lags that can be measured experimentally. As opposed to previous quantitative models \cite{Kahana2002,Lewandowsky}, it is founded on a single computational principle that has a clear functional meaning; namely, we assume that the system tries to maintain important memories at the expense of less important ones, and to this end, each newly acquired element erases already stored ones that are less significant. Importance is evaluated by a multi-dimensional valence measure such that memories that remain are characterized by relatively higher valence measures in one or more dimensions. The nature of these differential valence dimensions is not specified in the model. For example, the memory of an event could have five domains (who, when, where, what and why), each of them defining a different axis of importance. If that event involves a very relevant person (therefore a high value on the 'who' axis), it would be likely to be retained in memory, even if what happened was relatively insignificant. Another way to view a piece of information as a multidimensional element, comes from the work on semantic representations of words. In particular, it has been shown that the same word pertaining to different conceptual groups, activates different parts of the brain according to the contextual associations made upon acquisition \cite{Huth2016}. 

Similar to \cite{Kahana2002}, the average life-span of memories in the model diverges due to accumulation of very strong memories, and hence the process never reaches a steady state, with the number of memories increasing, albeit with monotonically decreasing speed. Besides the number of dimensions, the model does not have a single free parameter, hence the observation that it fits the experimental results so well is quite surprising. It shows that retroactive interference, which is well documented in psychological studies \cite{Wixted2004}, is by itself sufficient to account for power-law retention curves. Consolidation is thus not critical for this property of memory, which does not preclude its role in other aspects of memory that were not addressed in this study.

Some of the assumptions of the model clearly oversimplify the memory system (some of them are mentioned in Introduction). In particular, the valence of each memory is supposed to be stable for the duration of memory, and the distribution of valences, while not constrained in the model, is supposed to be stationary. In real life, one could imagine that some memories' importance could be altered in time while the distribution of new memory valences could also potentially change, for example due to aging or other life changes. It would be interesting to consider how the system would adapt to these changes by slowly replacing memories that become less relevant by the more relevant ones. 

Finally, we believe that we proposed a model with a minimal set of assumptions that is compatible with power-law forgetting curves. However, since other models, based on different principles, can also account for the same observations, we want to conclude the paper by encouraging further experimental and theoretical studies, in particular designing 'critical experiments', that would try to disambiguate different models and uncover the true mechanisms of forgetting. 

\section*{Methods}

\subsection*{Participants, Stimuli and Procedure}
A total of 900  participants were recruited to undertake a series of recognition tasks, designed to be performed utilizing Amazon's Mechanical Turk$^\circledR$ (mTurk) platform (https://www.mturk.com). Ethics approval was obtained by the IRB(Institutional Review Board) of the Weizmann Institute of Science and each participant accepted an informed consent form before participation. Participants were first required to complete the qualification task and if they met criteria described below they were allowed to participate in the main experiment (471 people). Participation was compensated at 10 cents for the qualification task and 30 cents for the regular task.

\paragraph{Delayed Recognition Tasks} All tasks performed in this study were two alternative forced choice (2AFC) delayed recognition tasks. Experiments were initiated with participants clicking on a 'Start Experiment' button. A stream of words was presented sequentially utilizing the standard interface on mTurk's website for Human Intelligence Tasks, using a custom HTML file with embedded Javascript. Each word was briefly flashed for a duration of $1s$ followed by a blank screen of $0.5s$. The words were displayed centrally on a white screen in black font. At random points during the trial and once in the end, after all words were presented, the presentation of words paused and participants were given a choice of two words in the form of vertically aligned buttons. Each button was randomly assigned with a word, one that was previously presented during the trial and one new word (lure). Participants were instructed to select the button containing the word they remembered seeing. After the selection, presentation resumed automatically. The list of presented words for each participant was randomly generated by sampling without replacement from a pool of 751 words which was produced by selecting English words\cite{Healey2014} that exhibited a frequency larger than ten per million\cite{Medler2005}. Each participant performed only one qualification and one main task trials.

\paragraph{Qualification} In previous experience we encountered that many workers on mTurk platform are not following instructions. Therefore, each participant was presented with simpler and shorter task first. A recognition delayed task with one hundred words in a stream was presented to participants. In 25 recognition tasks participants were questioned about word that was presented just before the last one (2-back task). We reasoned that two last presented words should stay in short-term memory if participants are attending to stimuli and following instruction. Therefore, we informed people who performed the qualification task with a success rate of more than 95\% that they may perform the main experiment. The rest were compensated for participation in qualification experiment.

\paragraph{Main Task} Similarly to the qualification task participants had to attend to a stream of words, in this case five hundred in total. During the trial, at seventy four random points (excluding the first 25 words) plus at the end of the list, they were prompted for a delayed recognition of a previously shown word versus a lure word. Twenty five of them requested a recognition of the second-back word as in the qualification, twenty five for the tenth-back and twenty five for the first twenty five words presented. Recognition tasks were randomly intermixed.

\subsection*{Analysis}
In Figure~\ref{fig:exp} the lag was computed as the difference between a query position and a presentation position in the stream of words. For example, if before the $100^{\mbox{th}}$ word there was a recognition task related to $15^{\mbox{th}}$ word the lag is $85$. In the figure the mean fraction of correct recognition is shown for lag bins with equal population of measurements (197) per bin, averaged across all participants having questions with query lags inside the bin. Not all participants had queries for all bins.

\subsubsection*{Correction for guessing}
In computing the theoretical performance for the recognition task we assumed that if a person is remembering the presented word then she/he would correctly point out to the presented word. In the case where participants do not remember the word, we assume they are guessing, and therefore choosing with equal probability. Therefore, one may express recognition performance as
\begin{equation}
    p(t) = R(t) + \frac{1}{2}(1-R(t)) = \frac{1+R(t)}{2}.
\end{equation}
 where $p(t)$ is fraction of correct responses in recognition task plotted in Figure~\ref{fig:avg_ret_filt} (dashed curves) and $R(t)$ is retention probability of a memory acquired $t$ time steps before testing.

\bibliographystyle{elsarticle-num}
\bibliography{sample}

\appendix
\section{Solution of Kahana model}
We analyze the version of Kahana model \cite{Kahana2002} with linear decay of memory strength: $S(t)=a-bt$ with positive random coefficients $a$ and $b$. Other types of passive decay produce similar results. For simplicity we assume that memory is forgotten when its strength dips below zero. The probability that a memory is still available time $t$ after its inception is given by 
\begin{equation}
    R(t)=Prob(a-bt>0)=\int_0^\infty db P_b(b) \int _{bt}^\infty da P_a(a)
\end{equation}
where $P_a$ and $P_b$ is the probability density of $a$ and $b$, respectively. Introducing the new variable $b \rightarrow bt $, and taking the limit $t \gg 1$, one obtains
\begin{align*}
    R(t)&=\frac{1}{t}\int_0^\infty db P_b(b/t) \int _{b}^\infty da P_a(a) \\
    &\approx \frac{P_b(0)}{t}  \int_0^\infty db \int _{b}^\infty da P_a(a) \\
    &=\frac{p_b(0)}{t}<a>
\end{align*}
where the third line is obtained by integration by part of the previous line, $<a>$ stands for the average value of $a$. Finally, we note that the $1/t$ scaling of retention implies that the probability density of the life-span of a memory, $P_{life}(t)$ scales as $1/t^2$ asymptotically for large $t$:
\begin{equation}
    P_{life}(t)=-\frac{d}{dt}R(t)\sim \frac{1}{t^2}
\end{equation}
which it turn implies that the average memory life-span is infinite.
\section{Multidimensional Retention Function Derivation}
\label{ap:1}
In the multidimensional case we can define the retention function as:
\begin{equation}
  R_n(t) =  \idotsint_{-\infty}^\infty P(v_1)P(v_2)\dotsi P(v_n)d(v_1)d(v_2)\dotsi d(v_n) \Big[1-\int_{v_1}^\infty \dotsi \int_{v_n}^\infty dv'_1 dv'_2 \dotsi dv'_n P(v'_1)P(v'_2)\dotsi P(v'_n)\Big]^t
\end{equation}
With a change of variable $v_i \rightarrow y(v_i)$ we get:
\begin{equation}
    R_n(t) = \idotsint_0^1 dy_1 \dotsi dy_n (1-y_1 \dotsi y_n)^t
\end{equation}
And computing the first integral:
\begin{align*}
    R_n(t) &= \idotsint_0^1 dy_1\dotsi dy_{n-1}  \left[\left. \frac{(1-y_1\dotsi y_n)^{t+1}}{(t+1) (y_1 \dotsi y_{n-1})}\right|_0^1  \right] \\
    &= \frac{1}{t+1}\int_0^1 dy_1\dotsi dy_{n-1} \left [\frac{(1-y_1\dotsi y_{n-1})^{t+1}-1}{(y_1 \dotsi y_{n-1})} \right] \\
    &= \frac{1}{t+1}\int_0^1 dy_1\dotsi dy_{n-1} \left[ \frac{1-(1-y_1\dotsi y_{n-1})^{t+1}}{1-(1-y_1 \dotsi y_{n-1})} \right]
\end{align*}
The term on the bracket is a progression and by expanding it we get:
\begin{equation}
    R_n(t) = \frac{1}{t+1}\idotsint_0^1 dy_1\dotsi dy_{n-1} \big[1+(1-y_1\dotsi y_{n-1})+(1-y_1\dotsi y_{n-1})^2 + \dotso + (1-y_1\dotsi y_{n-1})^t)\big]
\end{equation}
Rearranging terms:
\begin{equation}
    R_n(t) = \frac{1}{t+1} \Big[ \idotsint_0^1dy_1\dotsi dy_{n-1}(1-y_1\dotsi y_{n-1})^t + \idotsint_0^1dy_1\dotsi dy_{n-1}(1-y_1\dotsi y_{n-1})^{t-1} + \dotso + \idotsint_0^1dy_1\dotsi dy_{n-1}\Big]
\end{equation}
And by definition:
\begin{equation}
\label{eq:retention}
    R_n(t) = \frac{1}{t+1} \big[ R_{n-1}(t) + R_{n-1}(t-1) + \dotso + R_{n-1}(0) \big] = \frac{1}{t+1}\sum_{k=0}^t R_{n-1}(k)
\end{equation}
with $R_1(t)=\frac{1}{t+1}$ as derived in the text. In the limit of large $t$, one can replace the sums in equation \eqref{eq:retention} by corresponding integrals, which results in the approximate expression \eqref{eq:ret}. 
 
\section*{Acknowledgements}

This work is supported by the EU-H2020-FET 1564 and Foundation Adelis and EU - M-GATE 765549. We thank Michelangelo Naim for the help in designing and conducting Amazon Mechanical Turk\textregistered $ $ experiments.

\section*{Additional information}

\textbf{Competing interests} The authors have no competing interests or other interests that might be perceived to influence the results and/or discussion reported in this paper. 

\end{document}